\documentclass[aps,english,showpacs,twocolumn,superscriptaddress,pr]{revtex4-1}
\usepackage{amsfonts}
\usepackage{amssymb}
\usepackage{amsmath}
\usepackage{graphicx}
\usepackage{epsfig}
\usepackage{color}

\begin{document}

\title{High-order exceptional points in supersymmetric arrays}
\author{S. M. Zhang}
\affiliation{School of Physics, Nankai University, Tianjin 300071, China}
\author{X. Z. Zhang}
\affiliation{College of Physics and Materials Science, Tianjin Normal University, Tianjin
300387, China}
\author{L. Jin}
\email{jinliang@nankai.edu.cn}
\affiliation{School of Physics, Nankai University, Tianjin 300071, China}
\author{Z. Song}
\affiliation{School of Physics, Nankai University, Tianjin 300071, China}

\begin{abstract}
We employ the intertwining operator technique to synthesize a supersymmetric
(SUSY) array of arbitrary size $N$. The synthesized SUSY system is
equivalent to a spin-$(N-1)/2$ under an effective magnetic field. By
considering an additional imaginary magnetic field, we obtain a generalized
parity-time-symmetric non-Hermitian Hamiltonian that describes a SUSY array
of coupled resonators or waveguides under a gradient gain and loss; all the $%
N$ energy levels coalesce at an exceptional point (EP), forming the
isotropic high-order EP with $N$ states coalescence (EPN). Near the EPN, the
scaling exponent of phase rigidity for each eigenstate is $(N-1)/2 $; the
eigen frequency response to the perturbation $\epsilon$ acting on the
resonator or waveguide couplings is $\epsilon^{1/N} $. Our findings reveal
the importance of the intertwining operator technique for the spectral
engineering and exemplify the practical application in non-Hermitian physics.
\end{abstract}

\maketitle

\section{Introduction}

The exceptional point (EP) in a non-Hermitian system occurs when eigenstates
coalesce \cite{Bender,NMBook,AAlu}, and usually associates with the
non-Hermitian phase transition \cite{ChLu,ChenYF}. In a parity-time ($\mathcal{PT}$) symmetric non-Hermitian coupled system, the $%
\mathcal{PT}$ symmetry of eigenstates spontaneously breaks at the EP \cite%
{LFeng,AGuo,CE,BP,Hodaei,LFeng,SLonghi,LFeng17,Miri,Ozdemir,PTRev,JDu},
which determines the exact $\mathcal{PT}$-symmetric phase and the broken $%
\mathcal{PT}$-symmetric phase in this system. In the $\mathcal{PT}$%
-symmetric phase, the eigenvalues are real and the intensities oscillate as
a result of the nonorthogonality of eigenstates \cite{El}; in the broken $%
\mathcal{PT}$-symmetric phase, the intensities exponentially increase
because of the complex eigenvalues \cite{CE}. Besides the coupled
waveguide/resonator lattice, $\mathcal{PT}$-symmetric systems are simulated
by photonic quantum walks \cite{PXue1,PXue19,KWang}.

The EP has many applications in optics \cite%
{NMPRL08,Doppler,Xu,Assawaworrarit,Midya,HouZL,Alu}, not limited to
non-reciprocal energy transfer \cite{Xu}, unidirectional lasing \cite%
{PMiao16,Longhi17}, and optical sensing \cite{WChen,Hodaei17}. Bidirectional
lasing alters to unidirectional lasing when approaching the EP \cite%
{YangPNAS}; the direction of lasing is controllable through adjusting the
chiral mode of the micro resonator. Unidirectional lasing toward
single-direction is possible with gain and synthetic magnetic flux \cite%
{LJinPRL}. Moreover, the EP is a bifurcation point of the energy levels.
Near the EP, the eigen frequency response to the perturbation exhibits a
square-root dependence \cite{WChen} and a cubic-root \cite{Hodaei17}
dependence, respectively. In this regard, the EPs are useful for
sensing in comparison with the diabolic points; this feature has been
verified in optics, cavity optomechanics, cavity spintronics, and circuit
quantum electrodynamics \cite%
{Wiersig,ZPLiu,Shallem,Lau,MZhang,YHLai,MPH,Djorwe,Yan19,You19}. The
sensing susceptibility is greatly enhanced near the EPs \cite{CC}.

Different types of energy level coalescence exist in non-Hermitian systems.
The most common types of EPs are the two-state coalescence (EP2) that
exhibits a square-root dependence on the system parameters \cite%
{AGuo,CE,BP,WChen} and the three-state coalescence (EP3) that exhibits a
cubic-root dependence on the system parameters \cite%
{Eva,Hodaei17,Heiss,JQYou,HJing}. Even four-state coalescence (EP4) are
accessible in the coupled resonators \cite{KDing,JLEP4}. Recently, the
high-order EP of arbitrary order is realized in coupled resonators \cite%
{CTEPN}. And the scaling law for the eigenvalue and eigenstate confirm the
sensitive property of the high-order EP \cite%
{XPEP4,ZXZ,Eva08,LPan,CTChan19,Teimourpour}; the dynamics near a high-order
EP exhibits a power law dependence on the order of EPs for the maximal
amplification \cite{El-Ganainy18}.

The intertwining operator technique is a useful method for the spectral
engineering \cite{LonghiPRB,JLJPA}. Such technique is capable of eliminating
a target energy level in the spectral to create the isospectral SUSY partner \cite%
{ELSUSY,SCSUSY,NCSUSY,PRSUSY}, which has an identical spectrum except for the eliminated target level. In parallel, the intertwining
operator technique can add target energy level or realize Hamiltonian with
spectrum fully constituted by the desirable energy levels \cite{JLJPA}.
Exact solvable models with desirable energy levels can be synthesized
employing the intertwining operator technique. Thus, the intertwining
operator technique is beneficial for proposing non-Hermitian Hamiltonian
with multiple energy level coalescence. The concept of supersymmetry,
originated from quantum field theory \cite{SU}, has boomed during recent
years in the research fields of optics and photonics. It is possible to
create designed spectrum and propose intriguing applications using the
synthesized system. The SUSY array synthesized through intertwining operator
technique can be utilized for optical sensing~\cite{WChen,Hodaei17}, single
mode lasing \cite{ELSUSY,SCSUSY,PRSUSY}, and optical mode converting \cite%
{NCSUSY}. The integrated lasing array usually has multiple mode emission. To
acquire single mode lasing, one can design an isospectral partner SUSY array
using the intertwining operator technique. The spectrum of the partner SUSY
array is engineered to be constituted by all the excited-state mode except
for the ground-state mode of the lasing array. Coupling the partner SUSY
array to the lasing array and intentionally inducing loss in the partner
SUSY array can enable the ground-state single mode lasing \cite%
{ELSUSY,SCSUSY,PRSUSY}. The SUSY array can remove the ground-state mode of a
multimode light field and manipulate the modal content of the light field
through a hierarchical sequence of partner SUSY arrays \cite{NCSUSY}. The
synthetic SUSY is a hypercube useful for quantum information science \cite%
{Christandl}, and the proposed SUSY array has equally spaced energy levels;
thus the SUSY array is capable of realizing a perfect state transfer that
the initial state is exactly mapped from one side of the SUSY array to the
other side of it \cite{ZXZ}.

In this paper, we introduce the intertwining operator technique to propose a
non-Hermitian SUSY array of arbitrary size. The energy levels of the
proposed SUSY array are equally spaced square-root branches. The
non-Hermitian phase transition in the proposed SUSY array is associated with
an isotropic high-order EP. In contrast to the anisotropic
high-order EP \cite{CTChan19}, the isotropic high-order EP has identical parameter dependence for
independent system parameters in the parameter space. The Hamiltonian of the SUSY array can be
understood either as many uncorrelated spin-$1/2$ particles in a magnetic
field or the noninteracting bosonic many-particle system in a two-site
model. For the arbitrarily high-order EP, topological properties and the
frequency response to perturbation on the resonator couplings are
investigated. In the SUSY array of $N$ sites, the isotropic EPN has the
phase rigidity scaling exponent $\left( N-1\right) /2$. The eigen frequency
sharply responses to the coupling perturbation $\epsilon $ with the form $%
\epsilon ^{1/N}$ near the EPN. The results are in accord with topological
features of the EPN. The SUSY is proper for the perfect state transfer in
quantum information science.

The remainder of the paper is organized as follows. In Sec. \ref{II}, we
introduce the intertwining operator technique to synthesize the $\mathcal{PT}
$-symmetric SUSY array. In Sec. \ref{III}, we investigate the topological
properties of the arbitrarily high-order EP through the phase rigidity. In
Sec. \ref{IV}, we focus on the eigen frequency response to the coupling
perturbation to reflect features of the arbitrarily high-order EP. In Sec. %
\ref{V}, we summarize the results.

\section{$\mathcal{PT}$-symmetric SUSY array}

\label{II} 
\begin{figure}[t]
\includegraphics[bb=35 505 525 665, width=8.8 cm, clip]{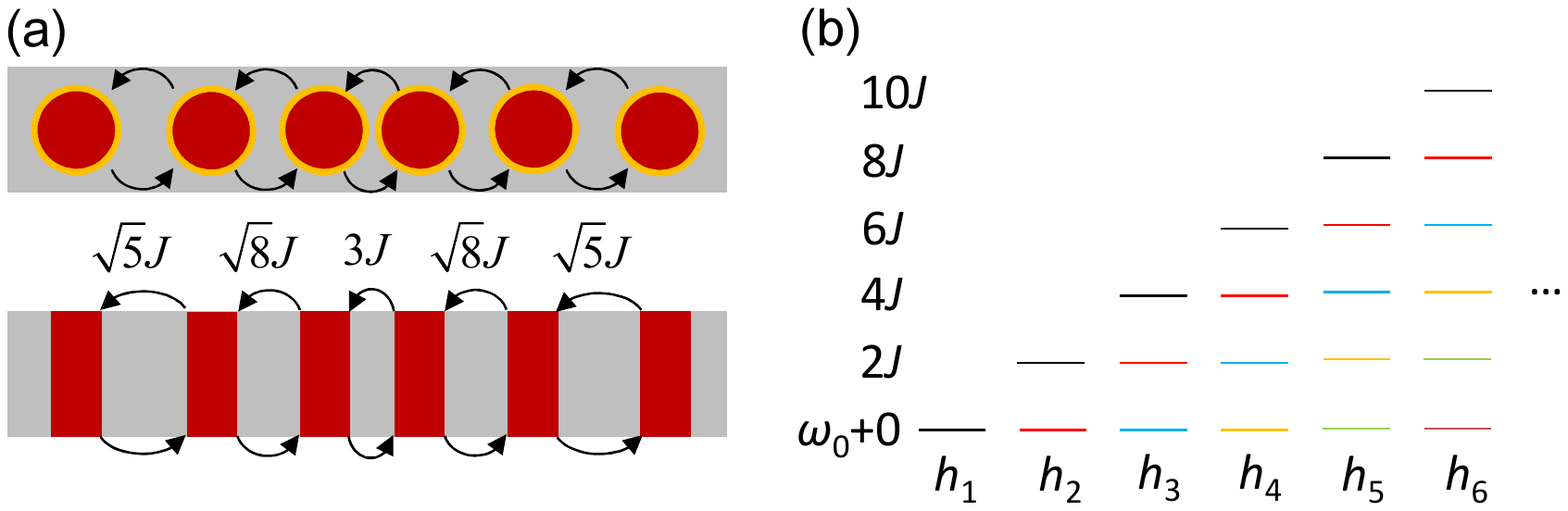}
\caption{(a) The SUSY array has a linear gradient on the gain
and loss distribution. The resonant frequency is $\protect\omega_0$ for each
resonator or waveguide. (b) Proposal of engineering the SUSY
array. } \label{fig1}
\end{figure}

In this section, we introduce the intertwining operator technique
to propose a synthetic SUSY array with all the energy levels equally spaced.
The synthetic SUSY is a hypercube, and the hypercube of arbitrary dimension
can be synthesized. The schematic of a synthetic six-site SUSY array of
coupled resonators (upper panel) or coupled waveguides (lower panel) is
shown in Fig. \ref{fig1}(a). The frequency of each resonator (waveguide) is $%
\omega _{0}$. The resonators (waveguides) are coupled through evanescent
tunneling between neighboring ones. The coupling amplitudes of the SUSY
array are determined from the intertwining operator technique. In Fig. \ref%
{fig1}(b), schematics of energy levels of the non-Hermitian SUSY arrays with
different site numbers are shown; all the levels are equally spaced and the
energy difference between each pair of neighboring energy levels is $2J$. To
constitute the SUSY array, the couplings are required to be engineered at
the proper amplitudes.

In the framework of quantum mechanics, the procedure is as follows. The
Hamiltonian $h_{N}$ has $N$ energy levels and we aim to remove an energy
level $\varepsilon _{N}$ from $h_{N}$ to construct a target superpartner $%
h_{N}^{\prime }$. The Hamiltonian $h_{N}$ is factorized into
\begin{equation}
h_{N}=Q_{N-1}R_{N-1}-\varepsilon _{N}I_{N},
\end{equation}%
where $Q_{N-1}$ is a $N\times \left( N-1\right) $ matrix; $R_{N-1}$ is a $%
\left( N-1\right) \times N$ matrix; and $I_{N}$\ is the $N\times N$ identity
matrix. The target superpartner Hamiltonian is obtained as
\begin{equation}
h_{N}^{\prime }=R_{N-1}Q_{N-1}-\varepsilon _{N}I_{N-1}.
\end{equation}%
$h_{N}^{\prime }$ is a $\left( N-1\right) \times \left( N-1\right) $ matrix
with $\left( N-1\right) $ levels identical to $h_{N}$ except for the energy
level $\varepsilon _{N}$.

To synthesize the SUSY array, we consider an inverse process and gradually
increase the size of the target Hamiltonian by adding target energy levels
one by one. Details of synthesizing the SUSY array are displayed as follows.
We start with a single-level system $h_{1}$, shift its energy level by $2J$
and add a zero energy ($\varepsilon _{2}=0$) to obtain energy spectrum $%
\{0,2J\}$ of the target Hamiltonian $h_{2}$. We factorize the Hamiltonian $%
h_{1}+2J=R$$_{1}Q_{1}$ with $R_{1}=\sqrt{J}(1,1)$ and take $Q_{1}$ as the
transpose of $R_{1}$, that is $Q_{1}=R_{1}^{T}$. The factorization is in the
form of
\begin{equation}
h_{1}+2J=R_{1}Q_{1}=\sqrt{J}\left(
\begin{array}{cc}
1 & 1%
\end{array}%
\right) \sqrt{J}\left(
\begin{array}{c}
1 \\
1%
\end{array}%
\right) =2J.
\end{equation}%
Then, following the intertwining operator technique, we interchange $R_{1}$
and $Q_{1}$ in the matrix product to obtain the target Hamiltonian
\begin{equation}
h_{2}=Q_{1}R_{1}=\sqrt{J}\left(
\begin{array}{c}
1 \\
1%
\end{array}%
\right) \sqrt{J}\left(
\begin{array}{cc}
1 & 1%
\end{array}%
\right) =J\left(
\begin{array}{cc}
1 & 1 \\
1 & 1%
\end{array}%
\right) .
\end{equation}%
The target Hamiltonian is a $2\times 2$ matrix $h_{2}=Q_{1}R_{1}=J\sigma
_{x}+JI_{2}$, where $\sigma _{x}$ is the Pauli matrix. Next, we offset the
energy by $-J$ to get $h_{2}-JI_{2}=J\sigma _{x}$ and gather the Hermitian
SUSY array of $N=2$. The matrix form of ${h}_{2}${\ after subtracting out
the term $JI_{2}$ is equivalent to the single-particle Hamiltonian for the
two-site model. }$J\sigma _{x}$ also describes a spin-$1/2$ particle in an
effective magnetic field $J$ along the $x$ direction, that is $\left(
B_{x},B_{y},B_{z}\right) =(J,0,0)$.

Furthermore, in order to construct the non-Hermitian Hamiltonian with the
EPs, we consider an additional effective imaginary magnetic field $i\gamma $
applied along the $z$ direction; the $\mathcal{PT}$-symmetric non-Hermitian
dimer is obtained with the expression $J\sigma _{x}+i\gamma \sigma _{z}$\
and the corresponding matrix form is given by
\begin{equation}
H_{2}=\left(
\begin{array}{cc}
\omega _{0}+i\gamma & J \\
J & \omega _{0}-i\gamma%
\end{array}%
\right) ,  \label{H1}
\end{equation}%
where $\omega _{0}$ is the on-resonator frequency of each resonator or
waveguide. $H_{2}$ has a pair of EP2s at $\gamma =\pm J$, where two
eigenstates coalesce to one $(\pm i,1)^{T}/\sqrt{2}$.

We repeat the above procedure to construct the non-Hermitian SUSY array with
high-order EPs. The spectrum of $h_{2}$ is shifted by $2J$ to $\left\{
2J,4J\right\} $; and then $\varepsilon_{3}=0$ is added to obtain the
spectrum $\{0,2J,4J\}$ of the target Hamiltonian $h_{3}$ [Fig. \ref{fig1}%
(b)]. We factorize $h_{2}+2J=R_{2}Q_{2}$ with $R_{2}=Q_{2}^{T}$, $%
Q_{2}\left( j,k\right) =\sqrt{J(j-1)}\delta _{j-1,k}+\sqrt{J(3-j)}\delta
_{j,j}$, where $\delta $\ means the Kronecker delta function. The
factorization gives%
\begin{equation}
h_{2}+2JI_{2}=R_{2}Q_{2}=J\left(
\begin{array}{cc}
3 & 1 \\
1 & 3%
\end{array}%
\right) .
\end{equation}%
Following the intertwining operator technique, we obtain a $3\times 3$
Hamiltonian after interchanging $R_{2}$ and $Q_{2}$ in the matrix product,
that is
\begin{equation}
h_{3}=Q_{2}R_{2}=J\left(
\begin{array}{ccc}
2 & \sqrt{2} & 0 \\
\sqrt{2} & 2 & \sqrt{2} \\
0 & \sqrt{2} & 2%
\end{array}%
\right) .
\end{equation}%
The target Hamiltonian $h_{3}$ can be expressed in the form of $%
h_{3}=Q_{2}R_{2}=JS_{x}+2JI_{3}$, where $S_{x(z)}$ is the angular momentum
operator for spin-$1$ in the $x$ ($z$) direction. We remove the overall
energy background $2JI_{3}$ from $h_{3}$ to obtain the Hermitian SUSY array
of $N=3$. $h_{3}-2JI_{3}=2JS_{x}$ describes a spin-$1$ particle in an
effective real magnetic field $(J,0,0)$.

By considering an effective imaginary field $i\gamma $ applied in the $z$
direction $(0,0,i\gamma )$, we obtain the $\mathcal{PT}$-symmetric
non-Hermitian trimer $JS_{x}+i\gamma S_{z}$. The corresponding non-Hermitian
SUSY array is
\begin{equation}
H_{3}=\left(
\begin{array}{ccc}
\omega _{0}+2i\gamma & \sqrt{2}J & 0 \\
\sqrt{2}J & \omega _{0} & \sqrt{2}J \\
0 & \sqrt{2}J & \omega _{0}-2i\gamma%
\end{array}%
\right) .  \label{H2}
\end{equation}%
The generalized SUSY array here is the two-particle Hamiltonian for the
two-site model in non-Hermitian cases, with the matrix form corresponding to
the Hamiltonian described by Eq.~(\ref{H2}). The quantum theory of spin
(angular momentum) tells us that the energy spectrum of the Hamiltonian $%
JS_{x}+i\gamma S_{z}$ is restricted to $n\sqrt{J^{2}-\gamma ^{2}}$ with $%
n=-2,0,2$. Each EP of $H_{3}$ belongs to EP3 with the eigen frequency $%
\omega _{0}$ and coalescence eigenstate $(-1,\pm i\sqrt{2},1)^{T}/2$ at the
critical condition $\gamma =\pm J$.

To synthesize the $4\times 4$ SUSY array, we shift all the energy levels of $%
h_{3}$ by $2J$ to obtain $\left\{ 2J,4J,6J\right\} $ and add a zero level to
compose the spectrum $\left\{ 0,2J,4J,6J\right\} $ of the target Hamiltonian
$h_{4}$. {We factorize }$h_{3}$ in the form of
\begin{equation}
h_{3}+2JI_{3}=R_{3}Q_{3}=J\left(
\begin{array}{ccc}
4 & \sqrt{2} & 0 \\
\sqrt{2} & 4 & \sqrt{2} \\
0 & \sqrt{2} & 4%
\end{array}%
\right) .
\end{equation}%
Following the intertwining operator technique, we obtain the $4\times 4$
target Hamiltonian%
\begin{equation}
h_{4}=Q_{3}R_{3}=J\left(
\begin{array}{cccc}
3 & \sqrt{3} & 0 & 0 \\
\sqrt{3} & 3 & 2 & 0 \\
0 & 2 & 3 & \sqrt{3} \\
0 & 0 & \sqrt{3} & 3%
\end{array}%
\right) .
\end{equation}%
We offset energy $3J$ from $h_{4}$ to obtain the Hermitian SUSY array of $%
N=4 $, which describes a spin-$3/2$ particle in an effective real magnetic
field $(J,0,0)$. The non-Hermitian generalization gives $JS_{x}+i\gamma
S_{z} $, where $S_{x(z)}$ is the angular momentum operator for spin $3/2$ in
the $x$ ($z$) direction. The matrix form of the SUSY array is
\begin{equation}
H_{4}=\left(
\begin{array}{cccc}
\omega _{0}+3i\gamma & \sqrt{3}J & 0 & 0 \\
\sqrt{3}J & \omega _{0}+i\gamma & 2J & 0 \\
0 & 2J & \omega _{0}-i\gamma & \sqrt{3}J \\
0 & 0 & \sqrt{3}J & \omega _{0}-3i\gamma%
\end{array}%
\right) .  \label{H3}
\end{equation}%
The $J$ related terms are the couplings between neighbor resonators induced
by the evanescent fields, and depend on the distance between them. The
distribution of gain and loss in the SUSY array has a gradient. The $%
\mathcal{PT}$-symmetric non-Hermitian $4\times 4$ SUSY array is the
three-particle Hamiltonian for the two-site model; the spectrum of which is
restricted to $n\sqrt{J^{2}-\gamma ^{2}}$ with $n=-3,-1,1,3$, with the
appearance of EP4 at $\gamma =\pm J$.

We can synthesize the SUSY array of arbitrary size via repeating the same
procedure. The $\mathcal{PT}$-symmetric non-Hermitian $5\times 5$ SUSY
array\ {has the form }%
\begin{equation}
H_{5}=\omega _{0}I_{5}+\left(
\begin{array}{ccccc}
4i\gamma & 2J & 0 & 0 & 0 \\
2J & 2i\gamma & \sqrt{6}J & 0 & 0 \\
0 & \sqrt{6}J & 0 & \sqrt{6}J & 0 \\
0 & 0 & \sqrt{6}J & -2i\gamma & 2J \\
0 & 0 & 0 & 2J & -4i\gamma%
\end{array}%
\right) .  \label{H4}
\end{equation}%
The $\mathcal{PT}$-symmetric non-Hermitian $6\times 6$ SUSY array
illustrated in Fig. \ref{fig1}(a) has the form
\begin{equation}
H_{6}=\omega _{0}I_{6}+\left(
\begin{array}{cccccc}
5i\gamma & \sqrt{5}J & 0 & 0 & 0 & 0 \\
\sqrt{5}J & 3i\gamma & \sqrt{8}J & 0 & 0 & 0 \\
0 & \sqrt{8}J & i\gamma & 3J & 0 & 0 \\
0 & 0 & 3J & -i\gamma & \sqrt{8}J & 0 \\
0 & 0 & 0 & \sqrt{8}J & -3i\gamma & \sqrt{5}J \\
0 & 0 & 0 & 0 & \sqrt{5}J & -5i\gamma%
\end{array}%
\right) .  \label{H5}
\end{equation}%
The SUSY arrays $H_{5}$ and $H_{6}$ are the four-particle and five-particle
Hamiltonians for the two-site model \cite{Eva08}, respectively. In both
cases, the EP occurs at $\gamma =\pm J$, but being EP5 in $H_{5}$ and EP6 in
$H_{6}$. Currently, experimental investigation on the high-order EP has
sprung up rapidly. For example, a fifth-order EP can be designed via tuning
parameters in nitrogen-vacancy centers \cite{PRREP5}; a $\mathcal{PT}$%
-symmetric electronic circuit has been proposed to study sensing at a
sixth-order EP \cite{XZEP6}.

In general, we shift all the energy levels of $h_{N-1}$ by $2J$ to obtain a
spectrum $\left\{ 2J,4J,\cdots ,2(N-1)J\right\} $ and add $\varepsilon
_{N}=0 $ by the intertwining operator technique to obtain $h_{N}$ with the
spectrum $\{0,2J,\cdots ,2(N-1)J\}$. We obtain $h_{N}=Q_{N-1}R_{N-1}=JS_{x}+%
\left( N-1\right) JI_{N}$ with $R_{N-1}=Q_{N-1}^{T}$, $Q_{N-1}\left(
j,k\right) =\sqrt{J(j-1)}\delta _{j-1,k}+\sqrt{J(N-j)}\delta _{j,j}$, and $%
S_{x(z)}$ is the angular momentum operator for spin $\left( N-1\right) /2$
in the $x$ ($z$) direction. Removing the offset energy $\left( N-1\right) J$
and introducing an imaginary magnetic field in the $z$ direction, we obtain
the $\mathcal{PT}$-symmetric non-Hermitian $N\times N$ model $JS_{x}+i\gamma
S_{z} $ \cite{ZXZ}. $H_{N}$ describes a $N$-site $\mathcal{PT}$-symmetric
non-Hermitian SUSY array \cite{NCSUSY,PRSUSY,SCSUSY}. The imaginary magnetic
field corresponds to tilted on-site imaginary potentials in the form of gain
and loss, which linearly depends on the site number. The concise form of $%
H_{N}$\ is given by
\begin{eqnarray}
H_{N} &=&\sum_{m=1}^{N-1}J\sqrt{m\left( N-m\right) }(\left\vert
m\right\rangle \left\langle m+1\right\vert +\mathrm{H.c.})  \notag \\
&&+\sum_{m=1}^{N}\left[ \omega _{0}+i\gamma \left( N+1-2m\right) \right]
\left\vert m\right\rangle \left\langle m\right\vert .  \label{HNM1}
\end{eqnarray}%
The non-Hermitian generalized SUSY array is synthesized by a recursive
bosonic quantization technique in the coupled resonators or waveguides \cite%
{Teimourpour}.

The first line in $H_{N}$ is the Hermitian SUSY array \cite{Christandl}. The
energy of $JS_{x}$ relates to the (quantized) possible value of spin angular
momentum in the $x$ direction, being $nJ$ for the integer $n=-\left(
N-1\right) ,-\left( N-3\right) ,\cdots ,\left( N-3\right) ,\left( N-1\right)
$. $H_{N}$ describes a spin-$(N-1)/2$ particle in an effective magnetic
field $\left( B_{x},B_{y},B_{z}\right) =\left( J,0,i\gamma \right) $. This
indicates that the SUSY array $H_{N}$ has the frequency
\begin{equation}
\omega _{N,n}=\omega _{0}+n\sqrt{J^{2}-\gamma ^{2}}.  \label{omega}
\end{equation}
Notably, the EPs of $H_{N}$ are exactly EPNs at $\gamma =\pm J$, where all
the levels are square-root branches and coalesce to the resonant frequency $%
\omega _{0}$.

Furthermore, introducing the angular momentum operators $S_{x}=a_{1}^{%
\dagger }a_{2}+a_{2}^{\dagger }a_{1}$, $S_{y}=ia_{2}^{\dagger
}a_{1}-ia_{1}^{\dagger }a_{2}$, and $S_{z}=a_{1}^{\dagger
}a_{1}-a_{2}^{\dagger }a_{2}$, the Hamiltonian $JS_{x}+i\gamma
S_{z}=J(a_{1}^{\dagger }a_{2}+a_{2}^{\dagger }a_{1})+i\gamma (a_{1}^{\dagger
}a_{1}-a_{2}^{\dagger }a_{2})\equiv H_{\mathrm{two-site}}$ can be
alternatively understood as the noninteracting bosonic many-particle system
in a $\mathcal{PT}$-symmetric non-Hermitian two-site model \cite{Eva08},
where $a_{1(2)}^{\dagger }$ and $a_{1(2)}$ are the creation and annihilation
operators of the first (second) site, respectively. $\left[ S_{a},S_{b}%
\right] =2i\epsilon _{abc}S_{c}$, where $\epsilon _{abc}$ is the Levi-Civita
symbol and $a,b,c$ are $x,y,z$. The commutation relation is equivalent to
that of Pauli matrices with spin $1/2$.

The basis set for the single particle system is chosen as $\left\vert
1\right\rangle _{1}=a_{1}^{\dagger }\left\vert \mathrm{vac}\right\rangle $, $%
\left\vert 2\right\rangle _{1}=a_{2}^{\dagger }\left\vert \mathrm{vac}%
\right\rangle $. $H_{2}$ in Eq.~(\ref{H1}) is $H_{\mathrm{two-site}}$ in the
single-particle basis. If we consider the two-particle problem, the basis
set is $\left\vert 1\right\rangle _{2}=(a_{1}^{\dagger })^{2}/\sqrt{2}%
\left\vert \mathrm{vac} \right\rangle $, $\left\vert 2\right\rangle
_{2}=a_{1}^{\dagger }a_{2}^{\dagger }\left\vert \mathrm{vac}\right\rangle $,
$\left\vert 3\right\rangle _{2}=(a_{2}^{\dagger })^{2}/\sqrt{2}\left\vert
\mathrm{vac} \right\rangle $. The factor $1/\sqrt{2}$ in the basis is not
only the normalization factor to ensure $_{2}\langle m|n\rangle _{2}=\delta
_{mn}$ ($m,n=1,2,3$); but also the normalization factor in the Fock
representation for the occupation number of two. $H_{3}$ in Eq. (\ref{H2})
is $H_{\mathrm{two-site} } $ in the two-particle basis. Moreover, the basis
set for three particle is $\left\vert 1\right\rangle _{3}=(a_{1}^{\dagger
})^{3}/\sqrt{6}\left\vert \mathrm{vac} \right\rangle $, $\left\vert
2\right\rangle _{3}=(a_{1}^{\dagger })^{2}a_{2}^{\dagger }/\sqrt{2}%
\left\vert \mathrm{vac}\right\rangle $, $\left\vert 3\right\rangle
_{3}=a_{1}^{\dagger }(a_{2}^{\dagger })^{2}/\sqrt{2 }\left\vert \mathrm{vac}%
\right\rangle $, $\left\vert 4\right\rangle _{3}=(a_{2}^{\dagger })^{3}/%
\sqrt{6}\left\vert \mathrm{vac}\right\rangle $. $H_{4}$ in Eq. (\ref{H3}) is
$H_{\mathrm{two-site}}$ in the three-particle basis. $H_{2}$ \cite{WChen}, $%
H_{3}$ \cite{Hodaei17}, and $H_{4}$ \cite{XPEP4} are experimentally realized
in different physical setups. In general, the basis for $\left( N-1\right) $%
-particle system is $\left\vert l\right\rangle _{N-1}=(a_{1}^{\dagger
})^{N-1-l}(a_{2}^{\dagger })^{l}/\sqrt{\left( N-1-l\right) !l!}\left\vert
\mathrm{vac}\right\rangle $, where $l=0,1,\cdots,N-1$ and the subscript in
the basis stands for the number of particles. $H_{ \mathrm{two-site}}$ in
the basis $\{\left\vert l\right\rangle _{N-1}\}$ gives $H_{N}$ in Eq. (\ref%
{HNM1}).

{The notion of SUSY plays an important role for plenty of intriguing optical
properties and functionalities as well as for a number of practical
applications of optical metamaterials \cite{ELSUSY,SCSUSY,NCSUSY,PRSUSY}. }%
The non-Hermitian SUSY array also provides a promising platform for the
study of the topology of arbitrary high-order EP. The geometric
topological properties reflect the order of EPs and are the essential
features of different EPs, which are captured by the phase rigidity scaling
exponents.

\begin{figure}[tb]
\includegraphics[bb=0 0 588 841, width=8.8 cm, clip]{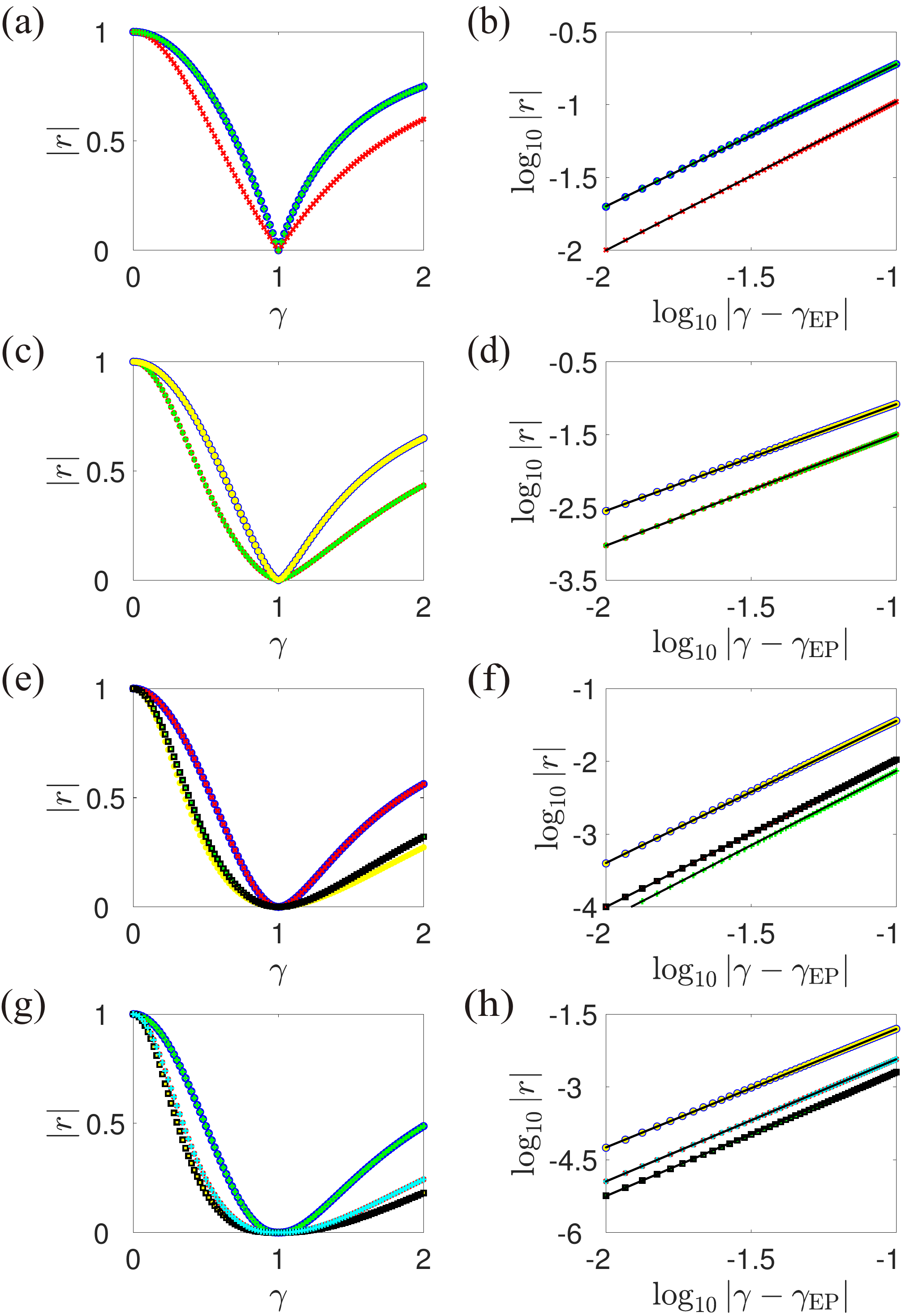}
\caption{Phase rigidities and scaling exponents of the high-order EP. (a), (c), (e), (g) and (b), (d), (f), (h) are the numerical
results of absolute values of the phase rigidities $\left\vert r\right\vert
=\left\vert \langle \protect\psi ^{\ast }|\protect\psi \rangle \right\vert
/\langle \protect\psi |\protect\psi \rangle $ {and the logarithmic
relationship between }$\left\vert r\right\vert $ and $\left\vert \protect\gamma -\protect\gamma _{\mathrm{EP}}\right\vert $ of energy levels at EP3, EP4, EP5 and
EP6, respectively. The phase rigidities are depicted in the left panel; the
scaling exponents in the right panel are $1$, $3/2$, $2$, and $5/2$ associated with EP3, EP4, EP5, and EP6. The black lines are the linear fits. The system
parameter is $J=1$ and all the EPs are at $\protect\gamma _{\mathrm{EP}}=1$.}
\label{fig2}
\end{figure}

\section{Phase rigidity}

\label{III} The phase rigidity is defined as
\begin{equation}
r=\left. \langle \psi ^{\ast }|\psi \rangle \right/ \langle \psi |\psi
\rangle ,
\end{equation}%
and $r$ reflects the mixing of different states $\psi ^{\ast }$ and $\psi $
\cite{Rotter,Eleuch}. We consider the phase rigidities associated with the
eigenstates of the SUSY array. At the EPs $\gamma _{\mathrm{EP}}=\pm J$, the
coalescence state is $\left\vert u_{1}\right\rangle =\left\vert
u_{2}\right\rangle =\left( \pm i,1\right) ^{T}/\sqrt{2}=\left\vert u_{%
\mathrm{EP}}\right\rangle $ for the non-Hermitian Hamiltonian $H_{2}$. The
eigen frequencies are $\omega _{0}+\varepsilon _{1}$ and $\omega
_{0}+\varepsilon _{2}$, where $\varepsilon _{1}=\sqrt{J^{2}-\gamma ^{2}}$, $%
\varepsilon _{2}=-\sqrt{J^{2}-\gamma ^{2}}$. The eigenstates satisfy $%
H_{2}\left\vert u_{j}\right\rangle =\left( \omega _{0}+\varepsilon
_{j}\right) \left\vert u_{j}\right\rangle $ ($j=1,2$) with $\left\vert
u_{1}\right\rangle =(e^{i\theta },1)^{T}/\sqrt{2}$ and $\left\vert
u_{2}\right\rangle =(-e^{-i\theta },1)^{T}/\sqrt{2}$, where $\cos \theta =%
\sqrt{J^{2}-\gamma ^{2}}/J\ $and $\sin \theta =\gamma /J$. $H_{N}$ at the
EPN has only one coalescence state for arbitrary $N$. Moreover, the
eigenstates of $H_{N}$ can be expressed as the direct product of $\left\vert
u_{1}\right\rangle $ and $\left\vert u_{2}\right\rangle $. Straightforward
calculation indicates $r_{\mathrm{EP}}=0$ for $H_{N}$ because of $\langle u_{%
\mathrm{EP}}^{\ast }|u_{\mathrm{EP}}\rangle =0$. The phase rigidity has a
scaling behavior near the EPs, $\left\vert r\right\vert =\left\vert \gamma
-\gamma _{\mathrm{EP}}\right\vert ^{\nu }$, where $\nu $ is the scaling
exponent and describes the\ topological feature of EPs.

The basis and eigenstates of the $\left( N-1\right) $-particle Hamiltonian
are chosen under the Fock representation. Considering the direct product
representation that employs the single-particle eigenstates as the basis,
the expression of eigenstates of the $\left( N-1\right) $-particle system is
the direct product of the $N-1$\ numbers of single-particle eigenstates $%
\left\vert u_{1}\right\rangle $\ and $\left\vert u_{2}\right\rangle $. The
general expression of the normalized eigenstate of a spin-$\left( N-1\right)
/2$\ system in the direct product representation is $\left( \left\vert
u_{1}\right\rangle \right) ^{l}\otimes \left( \left\vert u_{2}\right\rangle
\right) ^{N-1-l}$\ with the eigenvalue $l\varepsilon _{1}+\left(
N-1-l\right) \varepsilon _{2}$, where the integer $l\in \lbrack 0,N-1]$.
Although the expressions of Hamiltonians and eigenstates are formally
different under the two representations, the topological properties of EPs
remain unchanged. When approaching the EPN, the ratio of phase rigidities of
the eigenstates under the Fock representation to that under the direct
product representation is a constant, which does not affect the scaling law
near the EPN. The phase rigidity $r$ near the EPN under direct product
representation is $r=r_{1}^{l}r_{2}^{N-1-l}$; the scaling exponent $\nu $\
for EPN is $\nu =\log _{10}|r_{1}^{l}r_{2}^{N-1-l}|/\log _{10}|\gamma
-\gamma _{\text{\textrm{EP}}}|=\left( N-1\right) /2$ (see Appendix A). The
analysis on the scaling exponents of higher-order EPs is numerically
verified in Fig. \ref{fig2}, where the phase rigidities and scaling
exponents of EP3, EP4, EP5 and EP6 under the Fock representation Hamiltonian
[Eq. (\ref{HNM1})] are shown for $N=3,4,5,6$. The scaling exponents are $1.0$%
, $1.5$, $2.0$ and $2.5$ for the EP3, EP4, EP5, and EP6, respectively.
Notably, there exists a zero-energy flat band for odd $N$, which
participates in the coalescence of eigenstates (see Appendix B). Thus it
exhibits an identical scaling behavior to other levels. We emphasis that the
high-order EPs in the SUSY array are isotropic. Replacing $i\gamma $ by $%
\Delta +i\gamma $, we observe the scaling exponent $\nu =\left( N-1\right)
/2 $ for tuning the detuning $\Delta $ to approach the EPNs $\left\vert
r\right\vert =\left\vert \Delta -\Delta _{\mathrm{EP}}\right\vert ^{\nu }$.
The scaling exponent of the phase rigidity is robust to the perturbation
\cite{BBWei,CTChan19}. For an anisotropic EPN in the system with asymmetric
couplings, the scaling exponents of the phase rigidity are $(N-1)/2$ and $%
N-1 $ when approaching EPN from two independent parameters, respectively
\cite{CTChan19}.

\begin{figure}[tb]
\includegraphics[bb=0 0 600 651, width=8.8 cm, clip]{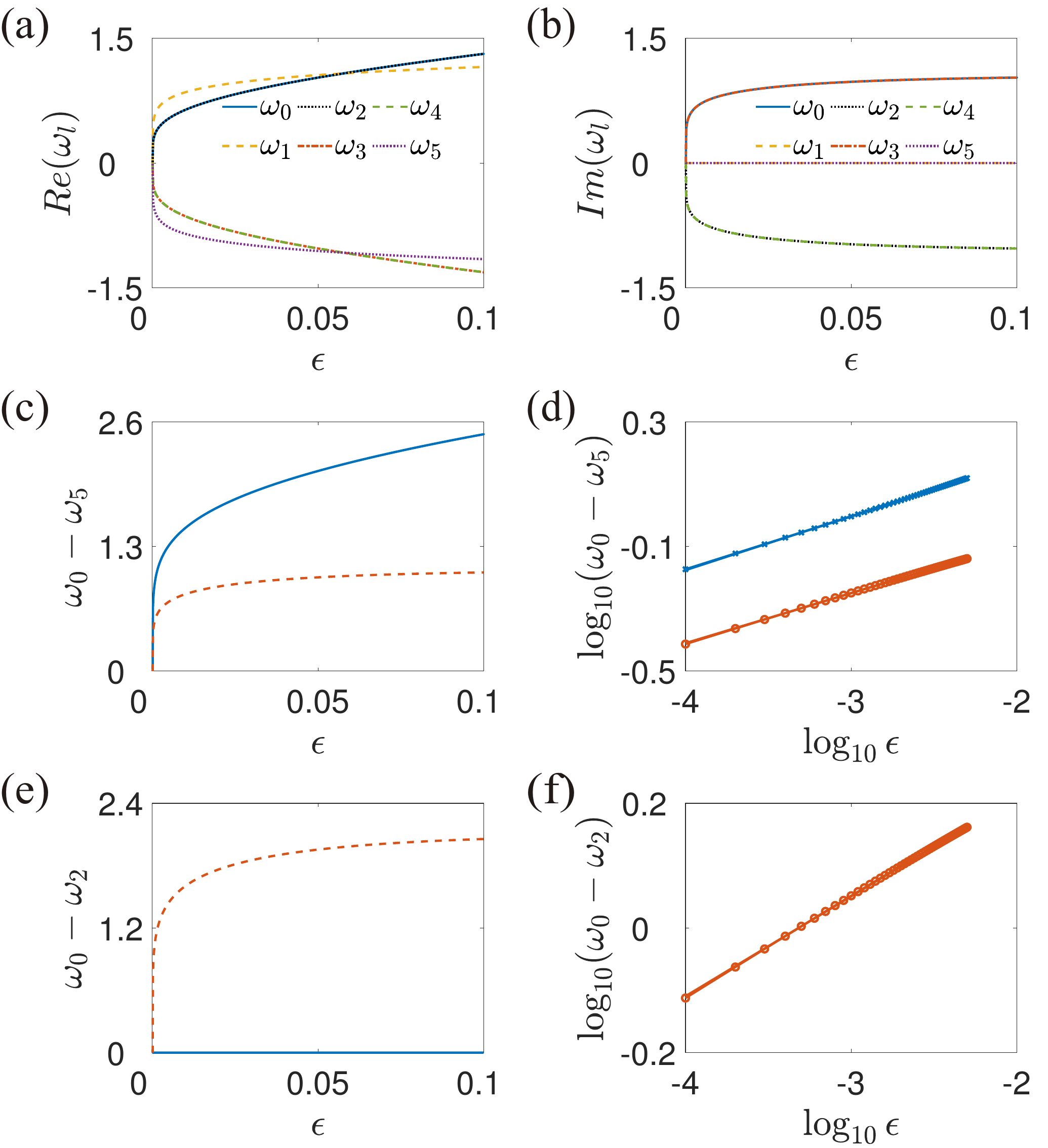}
\caption{Energy levels and frequency splittings near the EP6. (a) Real and (b)
imaginary parts of eigenvalues $\protect\omega _{l}$, $l=0,1,2,3,4,5$. (c)
and (e) are the frequency splittings between two bands $\protect\omega _{0}$,
$\protect\omega _{5}$ and $\protect\omega _{0}$, $\protect\omega _{2}$. (d)
and (f) are the logarithmic scales of (c) and (e), respectively. In (c), (e)
and (d), (f), dashed orange (solid cyan) lines and orange circles (cyan
crosses) correspond to the imaginary (real) parts. The slopes are $1/6$ in
(d) and (f). The system parameters are $\protect\gamma =J=1$.} \label{fig3}
\end{figure}

\section{Eigen frequency response to perturbation}

\label{IV}The non-Hermitian SUSY array at the high-order EP enhances the
susceptibility in optical sensing, the frequency response near the
high-order EP is greatly increased \cite{WChen,Hodaei17,CC,BBWei}. Near the
high-order EP in the non-Hermitian SUSY array, a remarkable point is the
enhanced frequency response to the detuning as well as the coupling when the
array is subjected to the perturbation $\epsilon $. The SUSY array is a
hypercube with high symmetry, the response to $\epsilon $ acting on the
coupling appears similar response to that acting on the detuning when
approaching an EPN. And the eigen frequency response is $\sim \epsilon ^{1/2}
$ for EP2 \cite{WChen,Djorwe} and $\sim \epsilon ^{1/3}$ for EP3 \cite%
{Hodaei17}, which is distinct from the linear response to the perturbation
strength $\sim \epsilon $ near the degeneracy point in Hermitian systems.
The sharp response is a typical feature of the EP that paves the way of the
application of sensors. Moreover, the SUSY array we constructed holds an
arbitrary EPN ($N\geq 2$) at $\gamma =\pm J$, which has striking features.

According to the Newton-Puiseux series expansion \cite{puiseux,puiseuxbook},
the frequency splitting $\omega _{l}$ is a function of the perturbation $%
c_{1}\epsilon ^{1/N}+c_{2}\epsilon ^{2/N}+\cdots $ \cite{BBWei}, where the
corresponding coefficients $c_{1}$, $c_{2}$, $\cdots $ are complex numbers.
In the perturbation theory, the unperturbed Hamiltonian is $H_{N}$, and the
perturbation Hamiltonian is $H^{\prime }$ ($H^{\prime }\ll H_{N}$). The SUSY
array under perturbations reads $H_{N\times N}=H_{N}+H^{\prime }$. The eigen
frequency of $H_{N\times N}$ and the expansion coefficients $c_{1}$ and $%
c_{2}$ are determined by the equation $\det \left[ H_{N\times N}-\omega
_{l}I_{N}\right] =0$ through substituting the first two terms $c_{1}\epsilon
^{1/N}+c_{2}\epsilon ^{2/N}$ of $\omega _{l}$. Because $\epsilon \ll J$, the
higher order terms of $\epsilon $ can be neglected. Notably, in general
cases, $\omega _{l}\approx c_{1}\epsilon ^{1/N} $ for comparable $c_{1}\sim
c_{2}$. We also show the eigen frequency response to the coupling
perturbation characterized by the order of magnitude $\sim \epsilon ^{1/N}$
near an EPN in the SUSY array is similar to that of the detuning
perturbation.

The coupling is determined by the distance between resonators or waveguides.
In practice, the resonator frequency can be accurately fabricated, and the
coupling $J$ may have imperfections. Thus, it is reasonable to consider the
resonator coupling perturbations for the EPN of the SUSY array $H_{N}$. We
take the example in Fig.~\ref{fig3} that all the coupling terms have
perturbations, that is, the $J$ related terms are $J\sqrt{ m\left(
N-m\right) }+\epsilon $ ($m=1,2,\cdots N-1$).

Figure \ref{fig3} depicts the energy levels, frequency splitting, as well as
the logarithmic plots of the frequency splitting as a function of coupling
perturbation $\epsilon $. The coupling perturbation $\epsilon $ presents in
each coupling $J\sqrt{m\left( N-m\right) }$. The whole spectrum is symmetric
about zero energy due to the equal amount of $\epsilon $ chosen without
breaking the chiral symmetry of $H_{N}$. The frequency splitting response
proportional to $\epsilon ^{1/N}$ is shown in Fig. \ref{fig3}; this differs
from the response to the coupling perturbation $J+\epsilon $ instead of $J$
in Eq. (\ref{omega}), which\ leads to the square-root dependence $\sim
\epsilon ^{1/2}$. The frequency splitting dependence on the coupling
perturbation between the first two cavities exhibits slopes of $1/3$ and $%
1/4 $ for $H_{3}$ and $H_{4}$ in the logarithmic plots; this reveals the
cubic-root and quartic-root dependence near the EP3\ and EP4 (see Appendix
C). In general cases, the coupling perturbation $\epsilon $ in an arbitrary
resonator of the SUSY array leads to the similar response $\omega
_{l}\approx c_{1}\epsilon ^{1/N}$; besides, we can observe the response is
in the order of $\omega _{l}\approx c_{2}\epsilon ^{2/N}$. The frequency
splitting response as a function of coupling perturbation is identical to
the detuning perturbation~\cite{LPan,Eva08}, which reflects the high
symmetry feature of the hypercube (SUSY array).

\section{Conclusion}

\label{V}The intertwining operator technique is an important approach for
the spectral engineering. We employ the intertwining operator technique to
propose the non-Hermitian SUSY array with arbitrarily high-order exceptional
points. The Hamiltonian of the proposed array is a non-Hermitian generalized
SUSY lattice chain for perfect state transfer in quantum information
science, which is equivalent to a noninteracting many-particle Hamiltonian
of the two-site non-Hermitian $\mathcal{PT}$-symmetric dimer. At the EPN of
the SUSY array with $N$ coupled resonators or waveguides, all the energy
levels are equally spaced, being square-root branches and coalescing at the
EPN. The phase rigidity of each eigenstate reaches zero and the scaling
exponent is $\left( N-1\right) /2$ for the EPN; the eigen frequency response
to perturbation $\epsilon $ is $\epsilon ^{1/N}$ for coupling amplitude
perturbation in certain resonators or waveguides of the SUSY array. The
intertwining operator technique provides a promising method for synthesizing
artificial optical metamaterial.

\section*{ACKNOWLEDGMENTS}

We acknowledge the support of the National Natural Science Foundation of
China (Grants No. 11975128, No. 11975166, and No. 11874225).

\section*{Appendix A: Phase rigidity}

The eigenvalues of $H_{2}$ are frequencies $\omega _{0}+\varepsilon _{1}$
and $\omega _{0}+\varepsilon _{2}$, where $\varepsilon _{1}=\sqrt{
J^{2}-\gamma ^{2}}$, $\varepsilon _{2}=-\sqrt{J^{2}-\gamma ^{2}}$. The
eigenstates satisfy $H_{2}\left\vert u_{j}\right\rangle =\left( \omega
_{0}+\varepsilon _{j}\right) \left\vert u_{j}\right\rangle $ with the
expressions $\left\vert u_{1}\right\rangle =(e^{i\theta },1)^{T}/\sqrt{2} $,
$\left\vert u_{2}\right\rangle =(-e^{-i\theta },1)^{T}/\sqrt{2}$, where $%
\cos \theta =\sqrt{J^{2}-\gamma ^{2}}/J\ $and $\sin \theta =\gamma /J$. We
start with the eigenstate $\left\vert u_{1}\right\rangle $\ of the $2\times
2 $ Hamiltonian $H_{2}$, where $J$ and $\gamma $ are positive real numbers
without loss of generality. At the EP2 $\gamma _{\text{\textrm{EP}} }=J$, $%
\left\vert u_{1}\right\rangle =\left( i,1\right) ^{T}/\sqrt{2}$, the phase
rigidity vanishes with $r_{\text{\textrm{EP}}}=0$.

For $J\geq \gamma $, the phase rigidity $r_{1}$ is
\begin{equation}
r_{1}=\left\vert \langle u_{1}^{\ast }\left\vert u_{1}\right\rangle /\langle
u_{1}\left\vert u_{1}\right\rangle \right\vert =\sqrt{1-\gamma ^{2}/J^{2}}.
\label{r1}
\end{equation}
The corresponding scaling exponent $\nu $ can be expressed as
\begin{equation}
\nu =\frac{\log _{10}\left\vert r\right\vert }{\log _{10}\left\vert \gamma
-\gamma _{\text{\textrm{EP}}}\right\vert }=\frac{1}{2}\lim_{\gamma
\rightarrow J}\frac{\log _{10}\left( \frac{J-\gamma }{J}\frac{J+\gamma }{J}
\right) }{\log _{10}\left( J-\gamma \right) }=\frac{1}{2}.
\end{equation}

For $\gamma \geq J$, the phase rigidity $r_{1}$ is
\begin{equation}
r_{1}=\left\vert \langle u_{1}^{\ast }\left\vert u_{1}\right\rangle /\langle
u_{1}\left\vert u_{1}\right\rangle \right\vert =\sqrt{1-J^{2}/\gamma ^{2}}.
\end{equation}
The corresponding scaling exponent $\nu $ can be expressed as
\begin{equation}
\nu =\frac{\log _{10}\left\vert r\right\vert }{\log _{10}\left\vert \gamma
-\gamma _{\text{\textrm{EP}}}\right\vert }=\frac{1}{2}\lim_{\gamma
\rightarrow J}\frac{\log _{10}\left( \frac{\gamma -J}{\gamma }\frac{\gamma
+J }{\gamma }\right) }{\log _{10}\left( \gamma -J\right) }=\frac{1}{2}.
\end{equation}
For the other eigenstate $\left\vert u_{2}\right\rangle $, we have the same
conclusion that $r_{2}=\sqrt{1-\gamma ^{2}/J^{2}}$ for $J\geq \gamma $\ and $%
r_{2}=\sqrt{1-J^{2}/\gamma ^{2}}$\ for $\gamma \geq J$. Therefore, the
scaling exponents\ are both $1/2$ for two eigenstates.

In general, the eigenstate for the $N\times N$ Hamiltonian $H_{N}$ with
eigenvalue $l\varepsilon _{1}+\left( N-1-l\right) \varepsilon _{2}$ is given
by $\left( \left\vert u_{1}\right\rangle \right) ^{l}\otimes \left(
\left\vert u_{2}\right\rangle \right) ^{N-1-l}$, where $l=0,1,\cdots ,N-1$.
Therefore, the phase rigidity vanishes with $r_{\text{\textrm{EP}}}=0$ at
EPN $\gamma _{\text{\textrm{EP}}}=J$ and the corresponding phase rigidity $r$
is
\begin{eqnarray}
r &=&\left\vert \frac{\langle \left( u_{1}^{\ast }\right) ^{l}\left(
u_{2}^{\ast }\right) ^{N-1-l}\left\vert u_{1}^{l}u_{2}^{N-1-l}\right\rangle
}{\langle u_{1}^{l}u_{2}^{N-1-l}\left\vert
u_{1}^{l}u_{2}^{N-1-l}\right\rangle }\right\vert  \notag \\
&=&\left\vert \frac{\langle \left( u_{1}^{\ast }\right) ^{l}\left\vert
u_{1}^{l}\right\rangle \langle \left( u_{2}^{\ast }\right)
^{N-1-l}\left\vert u_{2}^{N-1-l}\right\rangle }{\langle u_{1}^{l}\left\vert
u_{1}^{l}\right\rangle \langle u_{2}^{N-1-l}\left\vert
u_{2}^{N-1-l}\right\rangle }\right\vert  \notag \\
&=&r_{1}^{l}r_{2}^{N-1-l}.
\end{eqnarray}%
The scaling exponent $\nu $ for EPN is
\begin{eqnarray}
\nu &=&\frac{\log _{10}\left\vert r_{1}^{l}r_{2}^{N-1-l}\right\vert }{\log
_{10}\left\vert \gamma -\gamma _{\text{\textrm{EP}}}\right\vert }  \notag \\
&=&\frac{\log _{10}\left\vert r_{1}\right\vert ^{l}}{\log _{10}\left\vert
\gamma -\gamma _{\text{\textrm{EP}}}\right\vert }+\frac{\log _{10}\left\vert
r_{2}\right\vert ^{N-1-l}}{\log _{10}\left\vert \gamma -\gamma _{\text{
\textrm{EP}}}\right\vert }  \notag \\
&=&\frac{N-1}{2}.
\end{eqnarray}

\section*{Appendix B: Energy bands for odd $N$}

\begin{figure}[thb]
\includegraphics[bb=0 0 600 440,  width=8.8 cm, clip]{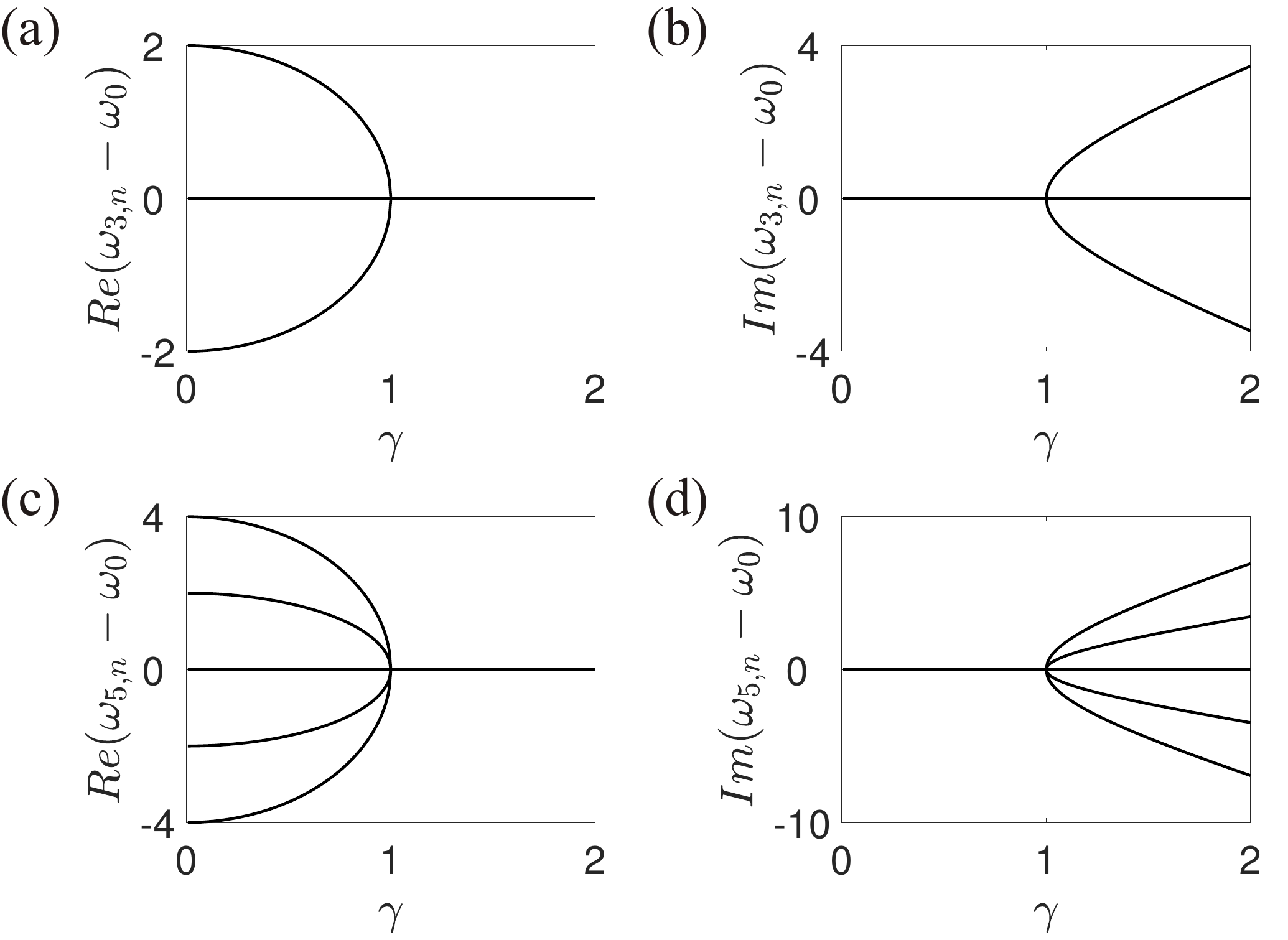}
\caption{Energy levels [Eq.~(\ref{omega})] of the SUSY array including a zero-energy level. (a) Real and
(b) imaginary parts of $H_{3}$. (c) Real and (d) imaginary parts of $H_{5}$. The system parameter is $J=1$.}
\label{fig4}
\end{figure}

A zero-energy flat band exists in $H_{N}$ when $N$ is odd because $H_{N}$
has the chiral symmetry; therefore, the spectrum of $H_{N} $ is symmetric
about the zero energy, and there is a zero-energy flat band if the system
has an odd number of energy levels. At the EP $\gamma =\pm J$, all energy
levels coalesce, $H_{N}$ is nondiagonalizable and reduces into a Jordan block%
\begin{equation}
\mathbf{D}_{N}=\left( d_{mn}\right) =\left\{
\begin{array}{l}
\omega _{0},m=n \\
1,m=n-1 \\
0,\mathrm{otherwise}%
\end{array}%
\right. .
\end{equation}

Figure \ref{fig4} shows the spectra of $H_{3}$ {[Eq. (\ref{H2})] }and $H_{5}
$ {[Eq. (\ref{H4})]} as a function of the non-Hermiticity $\gamma $. Both $%
H_{3}$ and $H_{5}$ hold a zero-energy band, which participates in the
coalescence at the EP $J=\gamma $.

\section*{Appendix C: Eigen frequency response to perturbation near the EP3,
EP4 and EP5}

\begin{figure}[thb]
\includegraphics[bb=0 0 600 440,  width=8.8 cm, clip]{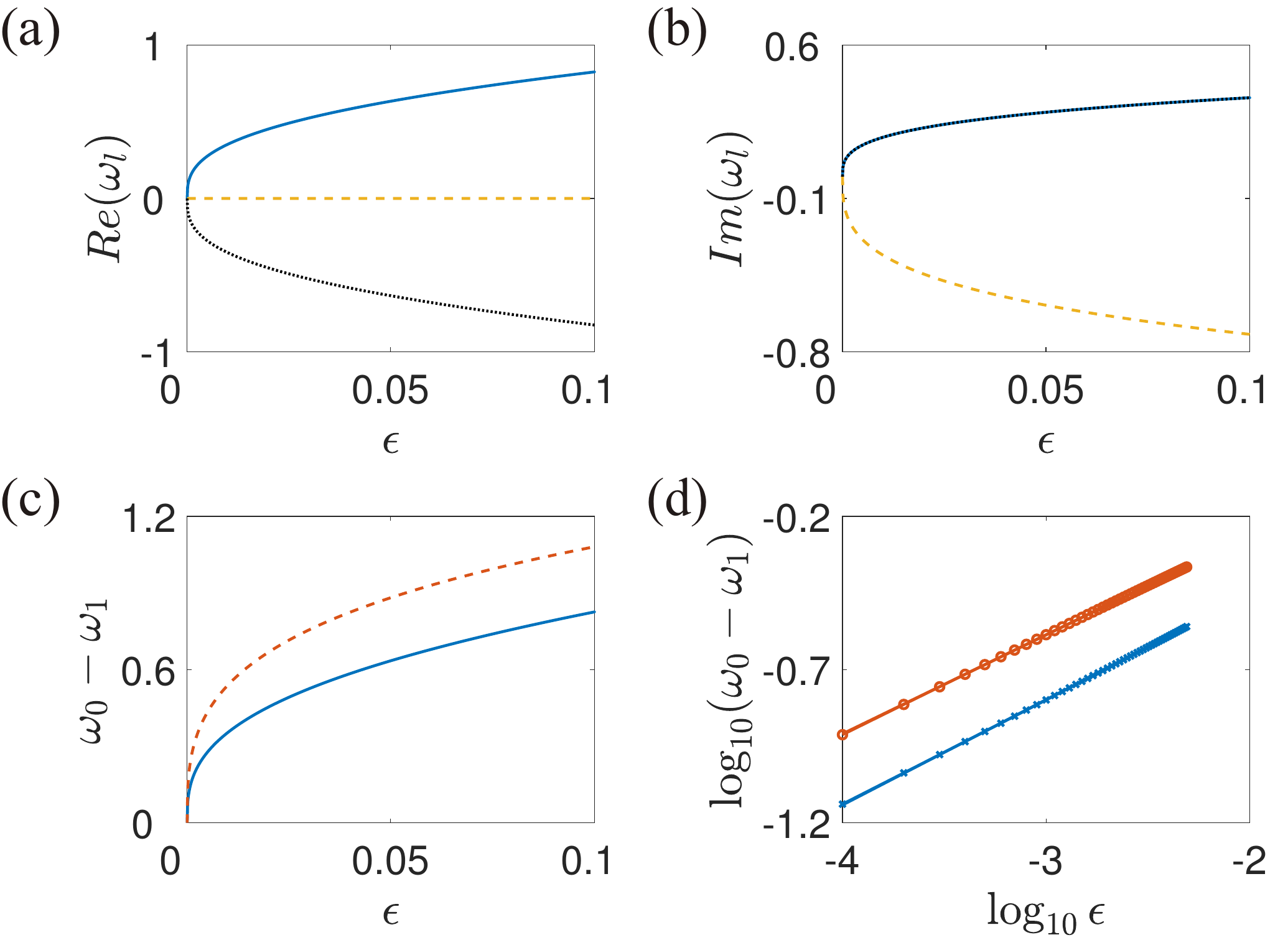}
\caption{Energy levels and frequency splittings near the EP3. (a) Real and (b)
imaginary parts of three eigenenergies with different colors and
line styles. (c) Frequency splitting between $\protect\omega _{0}$ and $\protect\omega _{1}$. (d) Results from (c) on a logarithmic scale. Cyan
(orange) color corresponds to the numerical results of the real (imaginary)
parts in (c) and (d). The slope is $1/3$ in (d). The system parameters are $\protect\gamma =J=1$.}
\label{fig5}
\end{figure}

Imposing perturbation $\epsilon $ on the couplings of the first two
resonators, we can get the matrix form for Hamiltonians $H_{3\times 3}$, $%
H_{4\times 4}$ and $H_{5\times 5}$,
\begin{equation}
H_{3\times 3}=\left(
\begin{array}{ccc}
\omega _{0}+2i\gamma & \sqrt{2}J+\epsilon & 0 \\
\sqrt{2}J+\epsilon & \omega _{0} & \sqrt{2}J \\
0 & \sqrt{2}J & \omega _{0}-2i\gamma%
\end{array}
\right) ,
\end{equation}
\begin{equation}
H_{4\times 4}=\left(
\begin{array}{cccc}
\omega _{0}+3i\gamma & \sqrt{3}J+\epsilon & 0 & 0 \\
\sqrt{3}J+\epsilon & \omega _{0}+i\gamma & 2J & 0 \\
0 & 2J & \omega _{0}-i\gamma & \sqrt{3}J \\
0 & 0 & \sqrt{3}J & \omega _{0}-3i\gamma%
\end{array}
\right) .
\end{equation}

\begin{equation}
H_{5\times 5}=\left(
\begin{array}{ccccc}
\omega _{0}+4i\gamma & 2J+\epsilon & 0 & 0 & 0 \\
2J+\epsilon & \omega _{0}+2i\gamma & \sqrt{6}J & 0 & 0 \\
0 & \sqrt{6}J & \omega _{0} & \sqrt{6}J & 0 \\
0 & 0 & \sqrt{6}J & \omega _{0}-2i\gamma & 2J \\
0 & 0 & 0 & 2J & \omega _{0}-4i\gamma%
\end{array}
\right) .
\end{equation}

Figure \ref{fig5}(c) represents the frequency splittings of bands with cyan
solid and yellow dashed lines in Fig.~\ref{fig5}(a) and Fig.~\ref{fig5} (b)
for $H_{3\times 3}$, the real and imaginary parts of which are depicted in
cyan solid and orange dashed lines. The logarithmic relationship between the
frequency splitting and disturbance is shown in Fig.~\ref{fig5}(d), where
the slope $1/3$ indicates the order of EP3.

\begin{figure}[tb]
\includegraphics[bb=0 0 600 440,  width=8.8 cm, clip]{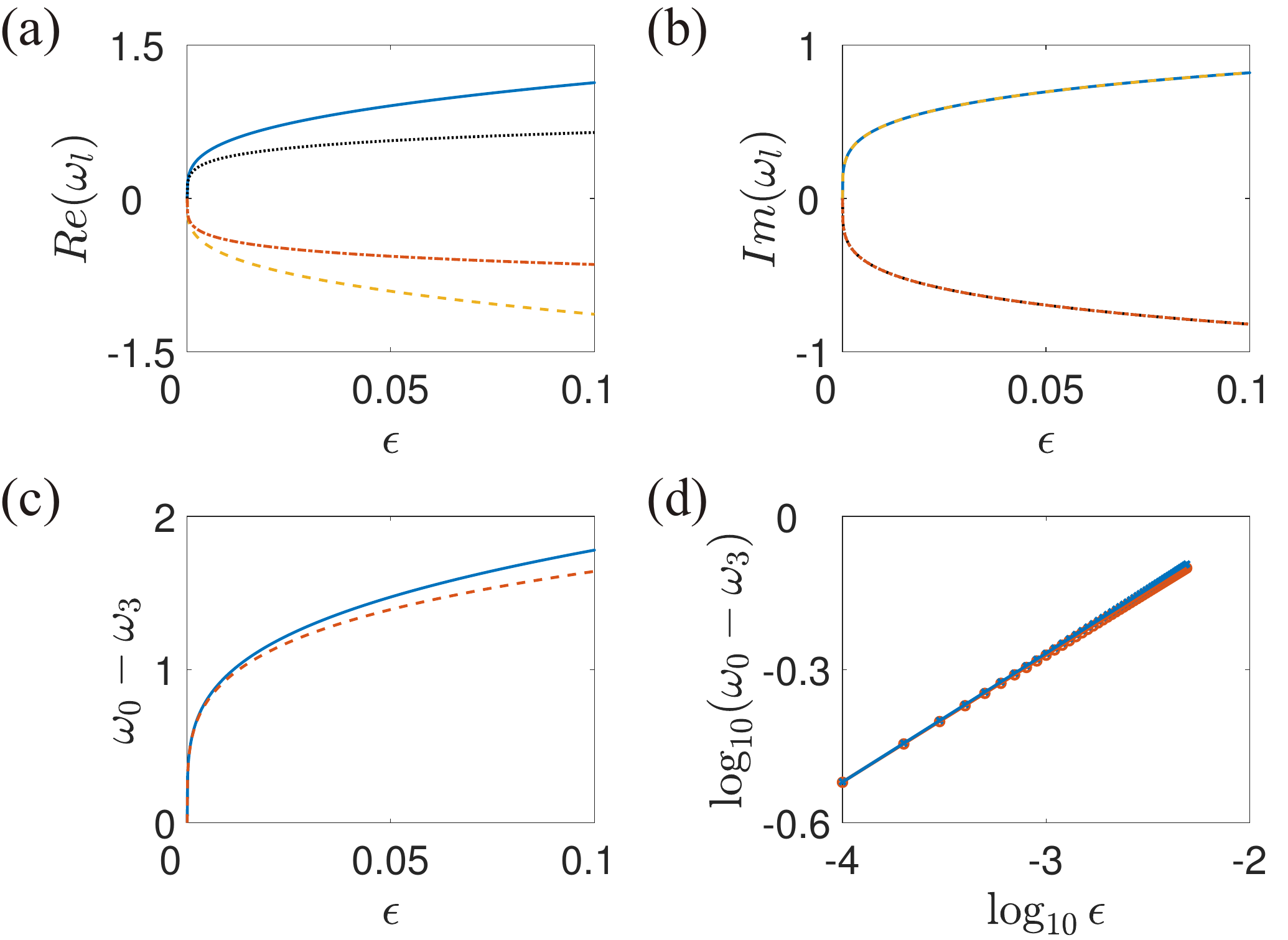}
\caption{Energy levels and frequency splittings near the EP4. (a) Real and (b)
imaginary parts of four eigenenergies with different colors and
line styles. (c) Frequency splitting between $\protect\omega _{0}$ and $\protect\omega _{3}$. (d) Results from (c) on a logarithmic scale. Cyan
(orange) color corresponds to numerical results of the real (imaginary)
parts of the SUSY array with $N=4$ in (c) and (d). The slope is $1/4$ in
(d). The system parameters are
$\protect\gamma =J=1$.} \label{fig6}
\end{figure}

\begin{figure}[tb]
\includegraphics[bb=0 0 600 440,  width=8.8 cm, clip]{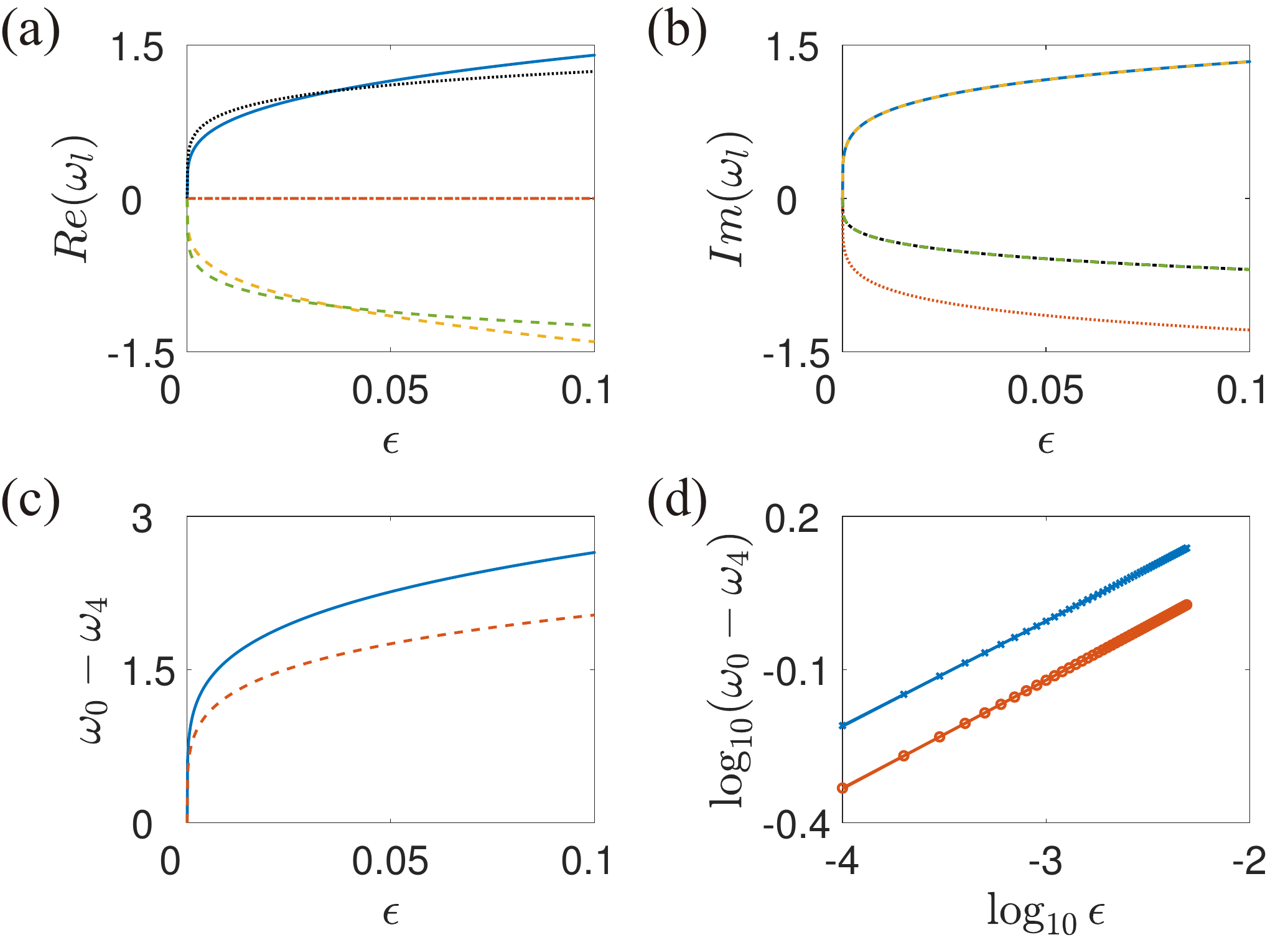}
\caption{Energy levels and frequency splittings near the EP5. (a)
Real and (b) imaginary parts of five eigenenergies with different colors and
line styles. (c) Frequency splitting between $\protect\omega _{0}$ and $\protect\omega _{4}$. (d) Results from (c) on a logarithmic scale. Cyan
(orange) color corresponds to numerical results of the real (imaginary)
parts in (c) and (d). The slope is $1/5$ in (d). The system parameters are $\protect\gamma =J=1$.}
\label{fig7}
\end{figure}

Two energy bands with the maximum and minimum real parts in Fig. \ref{fig6}%
(a) hold the identical positive imaginary parts in Fig. \ref{fig6}(b); the
middle two bands in Fig. \ref{fig6}(a) correspond to the same negative
imaginary parts in Fig. \ref{fig6}(b). Figure \ref{fig6}(c) represents the
frequency splittings of energy bands with cyan solid and orange dash-dot
lines in Fig. \ref{fig6}(a) and Fig. \ref{fig6}(b) for $H_{4\times 4}$, the
real (imaginary) part of which is depicted with cyan solid (orange dashed)
line. The corresponding logarithmic relationship with $\epsilon $ is shown
in Fig. \ref{fig6}(d), where the slope $1/4$ indicates the order of EP4.

Figure \ref{fig7}(c) represents the frequency splittings of energy bands
with cyan solid and green dashed lines in Fig.~\ref{fig7}(a) and Fig.~\ref%
{fig7}(b) for $H_{5\times 5}$, the real and imaginary parts of which are
depicted in cyan solid and orange dashed lines. The logarithmic relationship
between the frequency splitting and perturbation is shown in Fig.~\ref{fig7}%
(d), where the slope $1/5$ indicates the order of EP5.

\end{document}